\documentclass[%
reprint,
superscriptaddress,
showpacs,preprintnumbers,
amsmath,amssymb,
aps,
prx,
longbibliography
]{revtex4-2}

\usepackage{graphicx}
\usepackage{bm}
\usepackage{mathrsfs}
\usepackage{dsfont}
\usepackage{amsthm}
\usepackage{verbatim}
\usepackage{color}
\usepackage{bbold}
\usepackage{mathtools}
\usepackage[normalem]{ulem}
\usepackage[colorlinks,allcolors=blue]{hyperref}

\makeatletter

\newcommand{\Rmnum}[1]{\expandafter\@slowromancap\romannumeral #1@}
\newcommand{\mI}{{\mathcal I}}

\newcommand{\mS}{{\mathcal S}}

\newcommand{\mN}{{\mathcal N}}
\makeatother
\def\bra#1{\langle #1|}
\def\ket#1{ |#1 \rangle}

\def\Tr{\mathrm{Tr}}
\def\rank{\mathrm{rank}}
\def\mi{\mathrm{i}}

\begin{document}

\title{Experimental implementation of Hardy-like quantum pigeonhole paradoxes}

\author{Shihao Ru}
\affiliation{Ministry of Education Key Laboratory for Non-equilibrium Synthesis and Modulation of Condensed Matter, Shaanxi Key Laboratory of Quantum Information and Quantum Optoelectronic Devices, School of Physics of Xi'an Jiaotong University, Xi'an 710049, China}
\author{Cen-Xiao Huang}
\affiliation{CAS Key Laboratory of Quantum Information, University of Science and Technology of China, Hefei, 230026, China}
\affiliation{CAS Center For Excellence in Quantum Information and Quantum Physics, University of Science and Technology of China, Hefei, 230026, China}
\author{Xiao-Min Hu}
\affiliation{CAS Key Laboratory of Quantum Information, University of Science and Technology of China, Hefei, 230026, China}
\affiliation{CAS Center For Excellence in Quantum Information and Quantum Physics, University of Science and Technology of China, Hefei, 230026, China}
\author{Chao Zhang}
\affiliation{CAS Key Laboratory of Quantum Information, University of Science and Technology of China, Hefei, 230026, China}
\affiliation{CAS Center For Excellence in Quantum Information and Quantum Physics, University of Science and Technology of China, Hefei, 230026, China}
\author{Feiran Wang}
\affiliation{Ministry of Education Key Laboratory for Non-equilibrium Synthesis and Modulation of Condensed Matter, Shaanxi Key Laboratory of Quantum Information and Quantum Optoelectronic Devices, School of Physics of Xi'an Jiaotong University, Xi'an 710049, China}
\affiliation{School of Science of Xi’an Polytechnic University, Xi'an 710048, China}
\author{Ni Liu}
\affiliation{School of Mathematics and Statistics, Shaanxi Normal University, Xi'an 710119, China}
\affiliation{Yulin Gaoxin No.2 Middle School, Yulin 719000, China}
\author{Weidong Tang}
\email{wdtang@snnu.edu.cn}
\affiliation{School of Mathematics and Statistics, Shaanxi Normal University, Xi'an 710119, China}
\author{Pei Zhang}
\email{zhangpei@mail.ustc.edu.cn}
\affiliation{Ministry of Education Key Laboratory for Non-equilibrium Synthesis and Modulation of Condensed Matter, Shaanxi Key Laboratory of Quantum Information and Quantum Optoelectronic Devices, School of Physics of Xi'an Jiaotong University, Xi'an 710049, China}
\author{Bi-Heng Liu}
\email{bhliu@ustc.edu.cn}
\affiliation{CAS Key Laboratory of Quantum Information, University of Science and Technology of China, Hefei, 230026, China}
\affiliation{CAS Center For Excellence in Quantum Information and Quantum Physics, University of Science and Technology of China, Hefei, 230026, China}
\author{Fuli Li}
\email{flli@xjtu.edu.cn}
\affiliation{Ministry of Education Key Laboratory for Non-equilibrium Synthesis and Modulation of Condensed Matter, Shaanxi Key Laboratory of Quantum Information and Quantum Optoelectronic Devices, School of Physics of Xi'an Jiaotong University, Xi'an 710049, China}

\date{\today}
\begin{abstract}
We present the general Hardy-like quantum pigeonhole paradoxes for \textit{n}-particle states, and {find that each of such paradoxes can be simply associated to an un-colorable solution of a specific vertex-coloring problem induced from the projected-coloring graph (a kind of unconventional graph). Besides, as a special kind of Hardy's paradox,}
several kinds of { Hardy-like quantum pigeonhole} paradoxes can even give rise to higher success probability in demonstrating the conflict between quantum mechanics and local or noncontextual realism than the previous Hardy's paradoxes. {Moreover, not} only multi-qubit states, but high-dimensional states  can exhibit the paradoxes.
In contrast to only one type of contradiction presented in the original quantum pigeonhole paradox,
two kinds of three-qubit projected-coloring graph states as the minimal illustration are discussed in our work, and an optical experiment to verify such stronger paradox is performed.
This quantum paradox provides innovative thoughts and methods in exploring new types of stronger multi-party quantum nonlocality and may have potential applications in multi-party untrusted communications and device-independent random number generation.
\end{abstract}

\maketitle

\section{Introduction}\label{sec:Introduction}
Nowadays, using quantum approaches to solve some classical problems has attracted widespread attention.
A succession of new developed quantum technologies reveal their power in some specific tasks, such as quantum network \cite{kimble2008quantum,simon2017towards,wehner2018quantum}, fault-tolerant quantum computation \cite{GottesmanFTQC,2020TFTQC} and quantum metrology \cite{giovannetti2011advances}.
Based on the definite quantum feature, experts even proposed some quantum resource theories \cite{matera2016coherent,Hillery2016PRA,Theurer2017super,RMPresource}.
Quantum nonlocality \cite{Bell64,CHSH,GHZ89,Hardy93,cabello2000} and contextuality \cite{Specker1960,liu2016,hu2016,weidong2017,qu2021state,qi2022stronger} are two kinds of significant quantum resources. Not only they play a pivotal role in lots of quantum technologies, but also bring many counter-intuitive phenomena that may deep our understanding for the nature.
For example, the original quantum pigeonhole paradox \cite{aharonov2013,aharonov2016quantum,liu2020experimental,chen2019experimental,nmr2017pigenhole}, as a paradigmatic consequence of quantum nonlocality or contextuality \cite{yu2014quantum}, presents a counter-intuitive phenomenon: in a particular pre- and post-selection procedure \cite{Leifer2005,tollaksen2007pre}, if three particles (dubbed as three ``quantum pigeons'') are put into two boxes, then any pair of particles cannot stay in a same box, a violation to the classical pigeonhole counting principle \cite{aharonov2016quantum}.

Furthermore, quantum features can be visually exhibited by some specific mathematical structures including graphs \cite{TYO13,liu2021experimental}, braids \cite{Braid2005PRL}, knots \cite{hall2016tying}, as well as some new-defined geometric and topological objects \cite{chruscinski2012geometric,Kitaev2006}.
Performing coloring or sorting operations on such structures is a common thought to investigate the expected nonclassical features.
To finish such a task, generally, the main challenge resides how to accurately map the systems to the correlated mathematical objects, and how to make a suitable rule compatible with all the quantum relations but incompatible with one classical restriction at least.
A well-known example, the graphical proof of Kochen-Specker theorem \cite{KS67,Peres91,Cabello96} by the failure of noncontextuality (assigning values $0$ or $1$ to a set of rays), can be converted to a special coloring problem associated with the vertices of the induced orthogonal graph. Any pair of vertices connected by an edge indicates a mutually orthogonal relation of the correlated rays. In addition, we remind that only some special graphs can be regarded as the orthogonal graphs.

Recently, a special counter-intuitive mathematical object --- the projected-coloring graph (PCG), was presented in which, the nonclassical quantum features can be exhibited by the uncolorable PCGs \cite{Tang2021}.
In other words,  such uncolorable objects can only be simulated in some quantum processing rather than in the classical scenario. In fact, one can
even check that in some sense the minimal example of the uncolorable PCG can be considered as another description of Penrose triangle \cite{Penrose1958}.
Moreover, the purpose of introducing this mathematical object is to offer a more intuitive understanding for a new kind of quantum pigeonhole paradox called the ``Hardy-like quantum pigeonhole paradox''. 
Since the coloring problem in the PCGs belongs to a topological problem and (Hardy-like) quantum pigeonhole paradox is usually described in an algebraic framework, one can consider the former as a representation of the latter.
The Hardy-like quantum pigeonhole paradox exhibits the nonclassical counting principle resorting to a Hardy-like argument \cite{Hardy92,chen2018,luo2018experimental}.
The Hardy's proof is arguably considered as ``one of the strangest and most beautiful gems yet to be found in the extraordinary soil of quantum mechanics'' \cite{1994Mermin1}.
In addition, the Hardy-like paradoxes may have application in device-independent random number generation \cite{DIRNG_Han,hardy_randomness_amplication}.

In this work, first we give an introduction of the Hardy-like quantum pigeonhole paradox, and then show how to relate it with a classical vertex coloring problem of a kind of unconventional graph --- the projected-coloring graph. Here we realize a minimal experimental demonstration of this strong nonclassical feature by a heralded entanglement source of three entangled photons.
Compared with \cite{chen2018}, the demonstration of such Hardy-like quantum pigeonhole paradoxes does not require sophisticated techniques and are easier to implement. This quantum paradox may provide us a useful tool in exploring some new features of quantum mechanics and have some
useful applications in specific quantum information protocols such as multi-party quantum communication tasks.

\section{Theoretical description of Hardy-like quantum pigeonhole paradox}

\subsection{The simplest Hardy-like quantum pigeonhole paradox}
{Denote by $X_i~(Y_i,~Z_i)$ the usual Pauli matrix $\sigma_x~(\sigma_y,~\sigma_z)$ of the $i$-th qubit, and let
$|0\rangle_i,|1\rangle_i$ be two eigenstates of $Z_i$,  associated with eigenvalues $+1,-1$, respectively.}
Analogous to the conventional quantum pigeonhole paradox, the simplest system to show the Hardy-like quantum pigeonhole paradox requires three qubits \cite{Tang2021}, and the corresponding quantum state can be chosen as
\begin{subequations}
\begin{align}
\ket{S}&=\frac{1}{2}\left(\ket{000}-\ket{011}-\ket{101}-\ket{110}\right),\label{eq:S}\\
\ket{S^{\prime}}&=\frac{1}{2}\left(\ket{000}+\ket{011}+\ket{101}-\ket{110}\right),\label{eq:S'}
\end{align}
\end{subequations}
while $\ket{S}$ is discussed in this section and $\ket{S'}$ is in Appendix \ref{app:exp2}.
The state $\ket{S}$ is a special Greenberger-Horne-Zeilinger (GHZ) state ($\ket{S}=(\left|\circlearrowleft\circlearrowleft\circlearrowleft\right>+\left|\circlearrowright\circlearrowright\circlearrowright\right>)/\sqrt{2}$, where $|\circlearrowleft\rangle=(|0\rangle+i|1\rangle)/\sqrt{2}$ and $|\circlearrowright\rangle=(|0\rangle-i|1\rangle)/\sqrt{2}$).
Note that if the $i$-th qubit is measured and found in $\ket{0}$, the other two must be in the eigenstate of $X_jX_k$ with eigenvalue $-1$.  Therefore,
one can get the following quantum predictions,
\begin{subequations}
\begin{align}
&P(X_2X_3=-1|Z_1=+1)=1,\label{Hardy-Condi1}\\
&P(X_1X_3=-1|Z_2=+1)=1,\label{Hardy-Condi2}\\
&P(X_1X_2=-1|Z_3=+1)=1,\label{Hardy-Condi3}\\
&P(Z_1=Z_2=Z_3=+1)=\frac{1}{4}>0.\label{Hardy-Condi4}
\end{align}
\end{subequations}
Here $P(X_jX_k=-1|Z_i=+1)=1$ denotes the conditional probability that $X_j$ and $X_k$ are measured with outcomes
satisfying $X_jX_k =-1$ given the result of $Z_i = 1$. Besides, $P(Z_1= Z_2= Z_3 =+1)$ is the joint probability of obtaining $Z_1 = 1,Z_2 = 1,Z_3 = 1$.

Consider a run of the experiment that $Z_1,Z_2,Z_3$ are measured and the results $Z_1=1,Z_2=1,Z_3=1$ are obtained (the corresponding probability is $1/4$). Assume that the state $\ket{S}$ can be modeled by  a local realistic description.
Since $Z_1=1$ is obtained in this run of the experiment,  according to  Eq.(\ref{Hardy-Condi1}), if $X_2$ and $X_3$ were measured in this run,
their results should satisfy $X_2X_3=-1$, indicting that qubit 2 and 3 (``pigeons'') cannot stay in the same state(``box''). Likewise, from Eq.(\ref{Hardy-Condi2}) and Eq.(\ref{Hardy-Condi3}), we can infer that  qubits 1 and 3, qubits 1 and 2 cannot stay in the same box either,
contradicting with classical pigeonhole principle. We get a  three-qubit Hardy-like quantum pigeonhole paradox.

Besides, one can understand this paradox from the angle of conventional Hardy's paradox based on the  constraint $({I-X_2X_3})$$({I-X_1X_3})({I-X_1X_2})/8=0$ as well.
{Similar to the standard construction of the Hardy's paradox presented in Ref \cite{Hardy93}, by invoking four extra constraints $P(\frac{I-X_2X_3}{2}=1|Z_1=+1)=1,P(\frac{I-X_1X_3}{2}=1|Z_2=+1)=1,
P(\frac{I-X_1X_2}{2}=1|Z_3=+1)=1$, and $P(Z_1=Z_2=Z_3=+1)=\frac{1}{4}$, one can get a three-qubit common Hardy's paradox. We should stress that these constraints are more general in contrast with Eqs.(\ref{Hardy-Condi1})-(\ref{Hardy-Condi4}), since in the aforementioned argument of the Hardy-like pigeonhole paradox we have adopted a stronger  realistic assumption, namely, replacing $(X_iX_j)(\lambda)$ by $X_i(\lambda)X_j(\lambda)$, and therefore classical pigeonhole principle is required. In view of this,  the Hardy-like pigeonhole paradox is just a weaker version of the Hardy's paradox.}

Notice that in the above argument, we only discuss the contradiction arising from the case that $Z_1,Z_2,Z_3$ are measured with the results $Z_1=1,Z_2=1,Z_3=1$. In fact, by introducing other conditional probability constraints and considering other results
such as $Z_1=1,Z_2=-1,Z_3=-1$, one can construct three more similar paradoxes. Namely, sometimes a given quantum state may induce more than one Hardy-like quantum pigeonhole paradox. But for convenience, here we only discuss the paradox arising from the constraint given by the joint probability that all the involved $Z_i$ are measured with outcomes of $+1$.

\subsection{General Hardy-like quantum pigeonhole paradoxes}

The general Hardy-like quantum pigeonhole paradoxes can be constructed based on a special kind of states called PCG states. Let $\mN=\{1,2,\cdots,n\}$,
and a $n$-qubit PCG state can be defined as
\begin{align}
|P_n\rangle=\frac{1}{\sqrt{p+1}}(|00\cdots0\rangle-\sum_{{i\in \mI}}\theta_i|\vec{0}\rangle_{\overline{\mS}_i}|\vec{1}\rangle_{\mS_i}),\label{n-qubit-hardy-PH-key-state}
\end{align}
where $\mS_i\subset\mN$ and $\overline{\mS}_i=\mN-\mS_i$.
The size of $\mS_i$ satisfies $2\leq\left|\mS_i\right|<n$ and $|\mS_{i}\cup\mS_{j}|>\max\{|\mS_{i}|,|\mS_{j}|\}$ ($i\neq j$).
Besides,
$|\vec{0}\rangle_{\mS_i}\equiv\otimes_{k\in\mS_i}|0\rangle_k$, $|\vec{1}\rangle_{\mS_i}\equiv\otimes_{k\in\overline{\mS}_i}|1\rangle_k$,
$\theta_{i}=\pm1$ and $p=\sum_{i\in\mI}|\theta_{\mS_i}|$.
$\mI$ is an index set, which is used for labeling a group of specific subsets of $\mN$.
If the qubits in $\overline{\mS_i}$ are measured and found in $|\vec{0}\rangle_{\overline{\mS}_i}$, then the others are found in the eigenstate of $\prod_{k\in\mS_{i}}X_k$ with eigenvalue $\theta_{i}$.
Some tools are useful in the argument of Hardy-like quantum pigeonhole paradoxes. Firstly, a Hardy matrix $A$ with respect to a PCG state $|P_n\rangle$ can be defined as $A=\left(A_{ij}\right)_{|\mI|\times n}$ with elements $A_{ij}=1$ if $i\in \mI$ and $j\in \mS_i$, and $A_{ij}=0$ otherwise. Next, an argumented Hardy matrix $B$ is defined by a dilated matrix $B=\left(A{\Big|}\vec{\Theta}\right)$, where the $i$-th element of the $|\mI|\time1$ vector $\vec{\Theta}$ is $\Theta_i=({\theta_{\mS_{i}}+|\theta_{\mS_{i}}|})/{2}$. Then we can classify the PCG states into two families: (1) $\rank(A)=\rank(B)$; (2) $\rank(A)\neq \rank(B)$.

For a state $|P_n\rangle$ having the form of Eq. (\ref{n-qubit-hardy-PH-key-state}) and satisfying $\rank(A)\neq \rank(B)$,  we can get the following conditions:
\begin{align}\label{eq3}
&\prod_{k\in \mS_i}X_k=-\theta_i,~~(i\in\mI)\notag\\
&\mathrm{if}\ Z_{j_1}=Z_{j_2}=\cdots =Z_{j_{n-|\mS_i|}}=1,
\end{align}
where $j_1,j_2,\cdots,j_{n-|\mS_i|}\in\overline{\mS}_i$, and $Z_{j_l}=1$ $(l\in\{1,2,\cdots,n\})$ denotes an event of measuring $Z_{j_l}$ with an outcome of $+1$, and likewise $\prod_{k\in \mS_i}X_k=-\theta_i$. Eqs. (\ref{eq3}) have a total of $\left|\mI\right|$ items.
Besides, one has the following quantum prediction as well,
\begin{align}\label{n-qubit-Hardy-QH2}
  P(Z_{1}=Z_{2}=\cdots=Z_{n}=1)=\frac{1}{p+1}>0.
\end{align}
It is known from Ref. \cite{Tang2021} that $\rank(A)\neq \rank(B)$ will give rise to that $\prod_{k\in \mS_i}X_k=-\theta_i~(i\in\mI)$ cannot be simultaneously satisfied by any local hidden variable or noncontextual hidden variable models. Then the classical pigeonhole counting principle is broken, i.e., one get an $n$-qubit Hardy-like quantum pigeonhole paradox.
Based on Eqs. (\ref{n-qubit-hardy-PH-key-state}-\ref{n-qubit-Hardy-QH2}), we have designed three general kinds of PCG states, two kinds of general $n$-qubit states and one kind of $n$-qudit states (see in Appendix \ref{app_states}), which show the $n$-qubit and $n$-qudit paradoxes, respectively.

\subsection{Graph projected-coloring problems and beyond}

\begin{figure*}[t]
\centering
\includegraphics[width=0.7\linewidth]{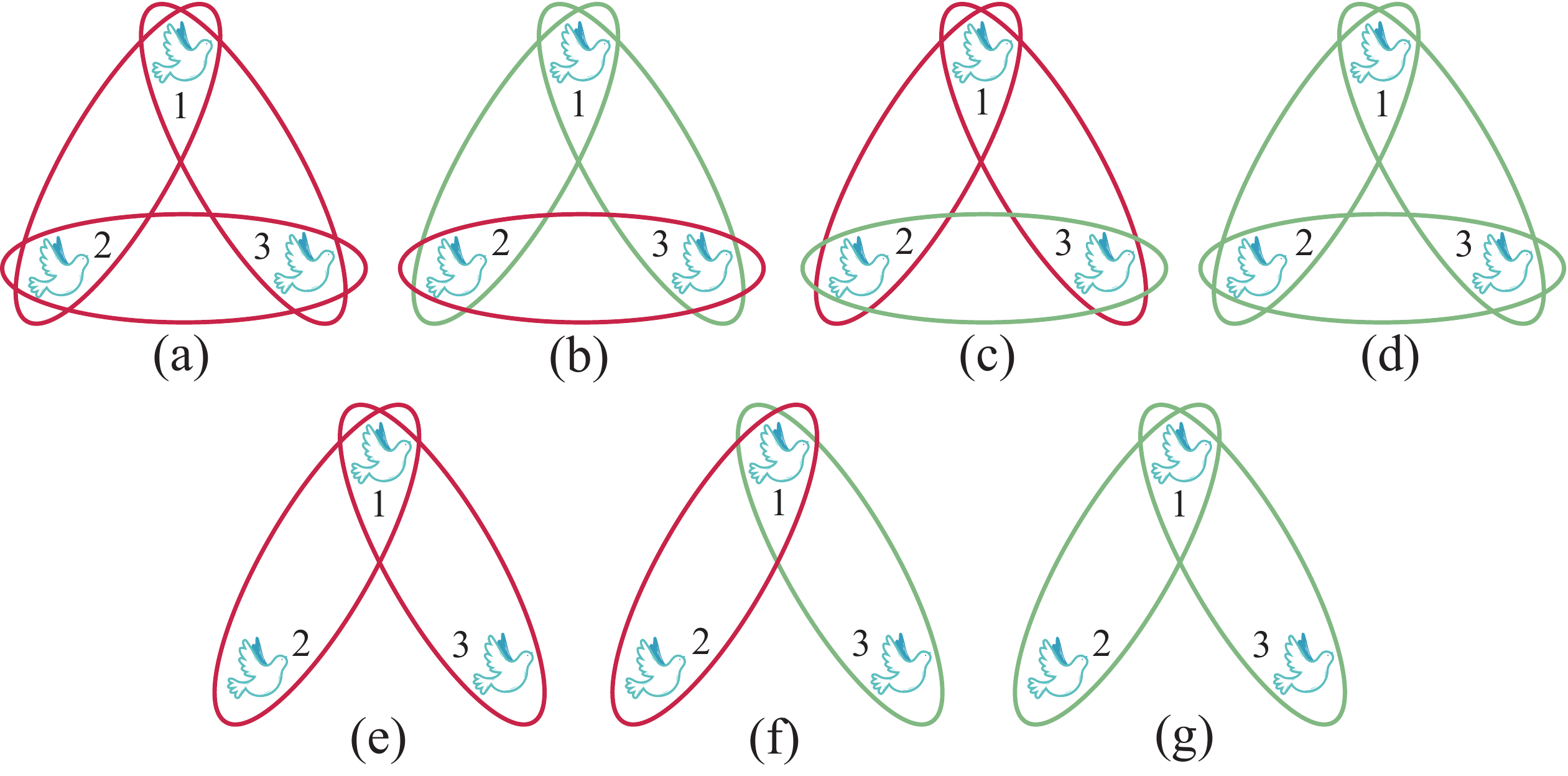}
\caption{(Color online). The graph projected-coloring problem studies the colorability of vertices under the condition of given edge colors.
There are seven categories of three-vertex PCGs, and the constraint relations of the edges in each category are equivalent.
The qubits on any edge can be regarded as a post-selected subsystem after projecting all the other qubits to the state $\ket{0}$. The color of this edge depends on the parity of the constraint which specifies the number of qubits staying in the box of $\ket{+}$.
A red edge ($\{j,k\}$) visually shows the constraint relation of different colors between the vertices $j$ and $k$.
It means that any pair of the qubits cannot stay in the same box ($X_jX_k=-1$) if the other qubit is post-selected to the state $|0\rangle$.
And a green edge ($\{j,k\}$) visually shows the constraint relation of the same color between the vertices $j$ and $k$.
(a) The first type of three-qubit uncolorable PCGs, which contains only one PCG, and corresponds to the PCG state $\left(\ket{000}-\ket{011}-\ket{101}-\ket{110}\right)/2$. (b) A representative example of  the second type of three-qubit uncolorable PCGs, which is associated with another one PCG state $\left(\ket{000}+\ket{011}+\ket{101}-\ket{110}\right)/2$. (c-g) Representative examples of all kinds of three-qubit colorable PCGs.}\label{3qubit-PCGs}
\end{figure*}

In contrast to the lack of pictorial representations for a common Hardy's paradox, one can find some interesting representations for the Hardy-like quantum pigeonhole paradox by some mathematical objects (e.g graphs) or problems.

First, we introduce that a vertex-coloring problem\cite{Tang2021} of an $n$-vertex PCG can be associated with the $n$-qubit Hardy-like quantum pigeonhole paradox, i.e., an $n$-vertex PCG $G$ to an $n$-qubit PCG state $|P_n\rangle$, where a $n$-vertex PCG can be defined as follows.

For any PCG state $|P_n\rangle$, a PCG $G$ is defined as an unconventional graph consisting of a set of vertices $V=\{1,2,\cdots,n\}$ and a set of weighted edges $E=\{\mS_i|i=1,2,\cdots,p\}$, where the weights of the edge $\mS_i$, say red($R$) and green($G$), correspond to the coefficient $\theta_{\mS_i}$ equals to $+1$ and $-1$, respectively.
In contrast to a usual graph, two or more (but less than $n$) vertices are allowed in an edge of an $n$-vertex PCG.
Moreover, to guarantee that there are no sub-edges inside any edge in $E$, an extra constraint should be imposed, namely, for any two edges $\mS_i$ and $\mS_j$, $|\mS_i\cup \mS_j|>\max\{|\mS_i|,|\mS_j|\}$.
Then the rule of graph projected-coloring problem can be described as follows:
\begin{itemize}
  \item[(1)] Each vertex $v_i\in V$ must be colored with either R or G, the coloring value $C(v_i)$ of a vertex $v_i$ is defined as $C(v_i)=-1$ if the vertex $v_i$ is colored with red, and $C(v_i)=+1$ otherwise;
  \item[(2)] The coloring value $C(\mS_i)$ of an edge $\mS_i$ can be defined as $C(\mS_i)=-1$ if the related weight is $R$, and $C(\mS_i)=+1$ otherwise;
  \item[(3)] A PCG $G$ is colorable if $\prod_{v_i\in\mS_i}C(v_i)=C(\mS_i)$ holds for any edge $\mS_i\in E$, otherwise uncolorable.
\end{itemize}

According to Ref. \cite{Tang2021}, any uncolorable PCG is associated with a proof of Hardy-like quantum pigeonhole paradox, which is related to a PCG state with $\rank(A)\neq \rank(B)$.
Note that the uncolorable PCG is a novel kind of impossible objects which cannot exist in the classical world.
To learn more about impossible objects, one can also check examples such as Penrose triangle \cite{Penrose1958} and Hilbert hotel\cite{HilbertHotel2015}. In fact, using impossible objects as analogies with some quantum features is not rare in the study of quantum theories, e.g. Hilbert hotel \cite{HilbertHotel2015}, and in this regard,  using the uncolorable PCG to illustrate the Hardy-like quantum pigeonhole paradox  can be considered as another example of such paradigm.

Take the case of three-vertex PCGs as an example.
As shown in Fig. \ref{3qubit-PCGs}, there are seven types of three-vertex PCGs, where Fig. \ref{3qubit-PCGs}(a) and \ref{3qubit-PCGs}(b) are uncolorable PCGs, and the others are all colorable PCGs.
As for Fig. \ref{3qubit-PCGs}(a) and (b), one can show three-qubit Hardy-like quantum pigeonhole paradox based on the state $\ket{S}$ and $\ket{S^{\prime}}$, respectively. We experimently measure the quantum feature based on these two states.
In all, $|S\rangle$ and $|S^{\prime}\rangle$ are two minimal states which can induce the Hardy-like quantum pigeonhole paradox.
They are local unitary equivalent as well as in experimental realization, but reveals different two kinds of the Hardy-like quantum pigeonhole paradox.
Moreover, in analogy to the hyper-graph state \cite{Gachechiladze2016}, here the PCG states can be considered as a kind of generalized graph states as well.

Besides the uncolorable PCG representation for the Hardy-like quantum pigeonhole paradox, one can find other representations for some special kind of such paradoxes, e.g. the quantum map coloring representation \cite{Tang2021}. Another interesting version for a kind of specific Hardy-like quantum pigeonhole paradox is the quantum magic square paradox (for more details, see in Appendix \ref{app_magic}).

Compared with other versions of Hardy's paradox, the Hardy-like quantum pigeonhole paradox has at least two advantages: (1) a simpler and more intuitive graphical representation (PCGs); (2) sometimes a higher success probability to show the contradiction between quantum mechanics and local or noncontextual realism. For example, consider the Hardy-like quantum pigeonhole paradox based on the PCG state $\ket{S_1(n)}$ (Appendix \ref{app_states}),
we can find that the success probability is $1/(n+1)$. By contrast, in the generalized $n$-qubit Hardy's paradox \cite{chen2018}, this probability can only be $1/2^{n-1}$.

\section{Experiment}
\subsection{Methods}
\begin{figure}[htbp]
  \centering
  \includegraphics[width=\linewidth]{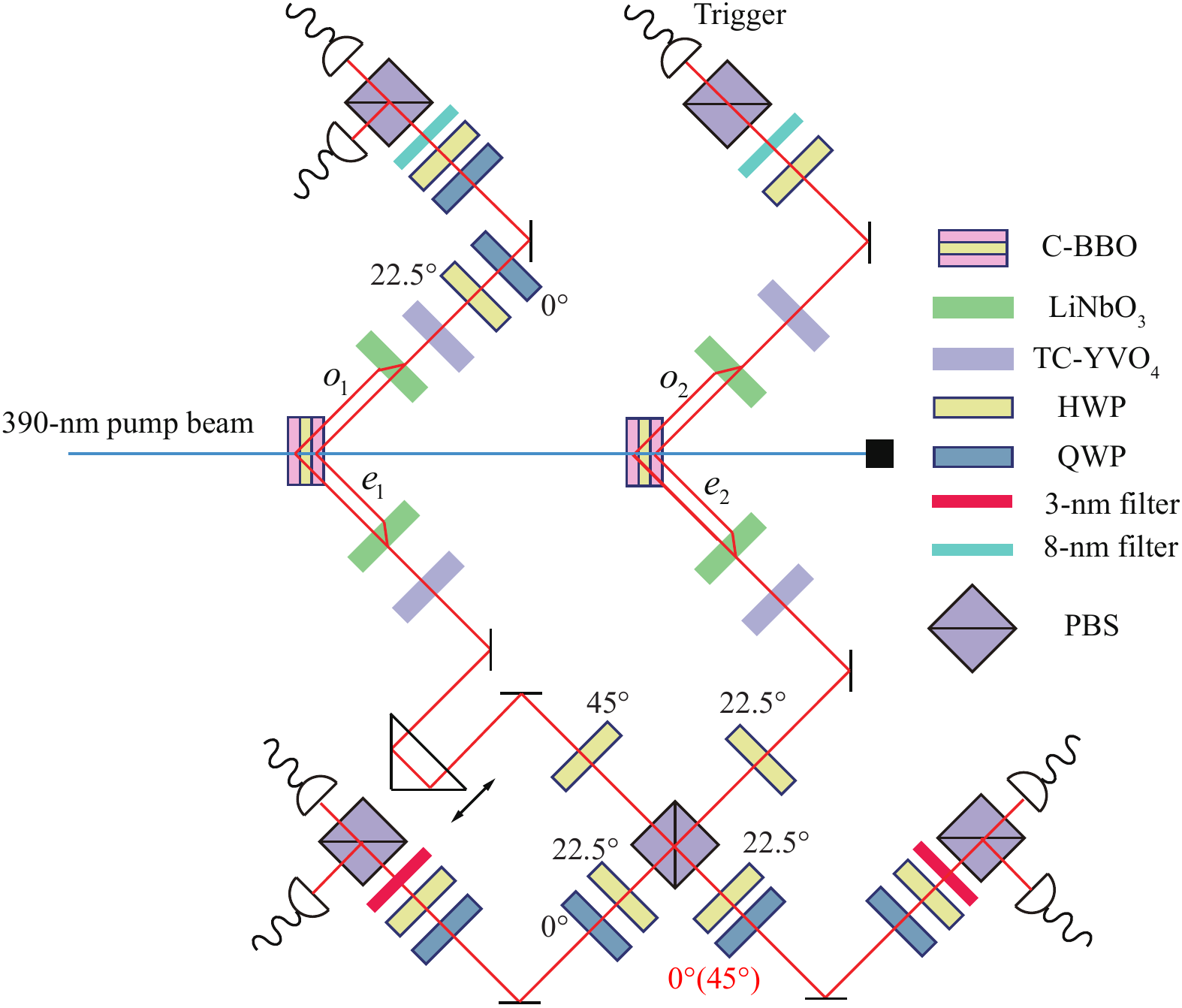}
  \caption{(Color online). Experimental setup. The pump beam, with a central wavelength 390 nm, a duration of 140 fs and a repetition rate of 76 MHz, passes through two c-BBO crystals successively to generate two pairs of polarized-entangled photons. Photons in spatial modes $e_1$ and $e_2$ overlap on PBS, a movable prime is used to guarantee the temporal delay between two paths are equal. Photons in modes $e_1$ and $e_2$ are spectrally filtered with 3-nm bandwidth filters, and photons in modes $o_1$ and $o_2$ are spectrally filtered with 8-nm bandwidth filters for a better collection efficiency.
We show the relative phase exists for two states $\ket{S}$ and $\ket{S^{\prime}}$, only changing one QWP's angle from $0^{\circ}$ to $45^{\circ}$ (red mark).
c-BBO: compound $\beta$-barium borate crystal; $\mathrm{LiNbO_3}$: $\mathrm{LiNbO_3}$ crystal for spatial compensation; $\mathrm{TC}$-$\mathrm{YVO_4}$: $\mathrm{YVO_4}$ crystal for temporal compensation; HWP: half-wave plate; QWP: quarter-half plate; PBS: polarizing beam splitter.}\label{fig:exp}
\end{figure}

Our experimental setup for generating the states $\ket{S}$ and $\ket{S^{\prime}}$ is illustrated in Fig. \ref{fig:exp}. We first prepare two pairs of polarized-entangled photons in the $\ket{\psi^-}=(\ket{H}\ket{V}-\ket{V}\ket{H})/\sqrt{2}$, in which $H(V)$ denotes $\ket{0}$ ($\ket{1}$) and the horizontal (vertical) polarization state of photons.
We adopt type-\Rmnum{2} phase-match beamlike compound $\beta$-barium (c-BBO) crystals, where a 780-nm true-zero-order half-wave plate (HWP) inserted between BBOs, to obtain polarized-entangled photons.

Pumped by ultraviolet laser pulses, photons pairs in the same state $\ket{H_e}\ket{V_o}$ are generated through spontaneous parametric down conversion since two BBO crystals have the same cutting angle and are placed in the same manner.
Here subscripts $o(e)$ indicate two spatial modes on which photons are ordinary (extraordinary) light.
Passing the true-zero-order HWP at $45^{\circ}$, the state of photon pairs generated by the left BBO is converted to $\ket{V_e}\ket{H_o}$.
By careful spatial and temporal compensations, polarized-entangled photons in the state $\ket{\psi^-}=(\ket{H_e}\ket{V_o}-\ket{V_e}\ket{H_o})/\sqrt{2}$ are prepared successfully.
More details about this sandwich-structure c-BBO for generating polarized-entangled source are in Ref. \cite{2016sixGHZ}.

\begin{figure}[t]
  \centering
  \includegraphics[width=0.8\linewidth]{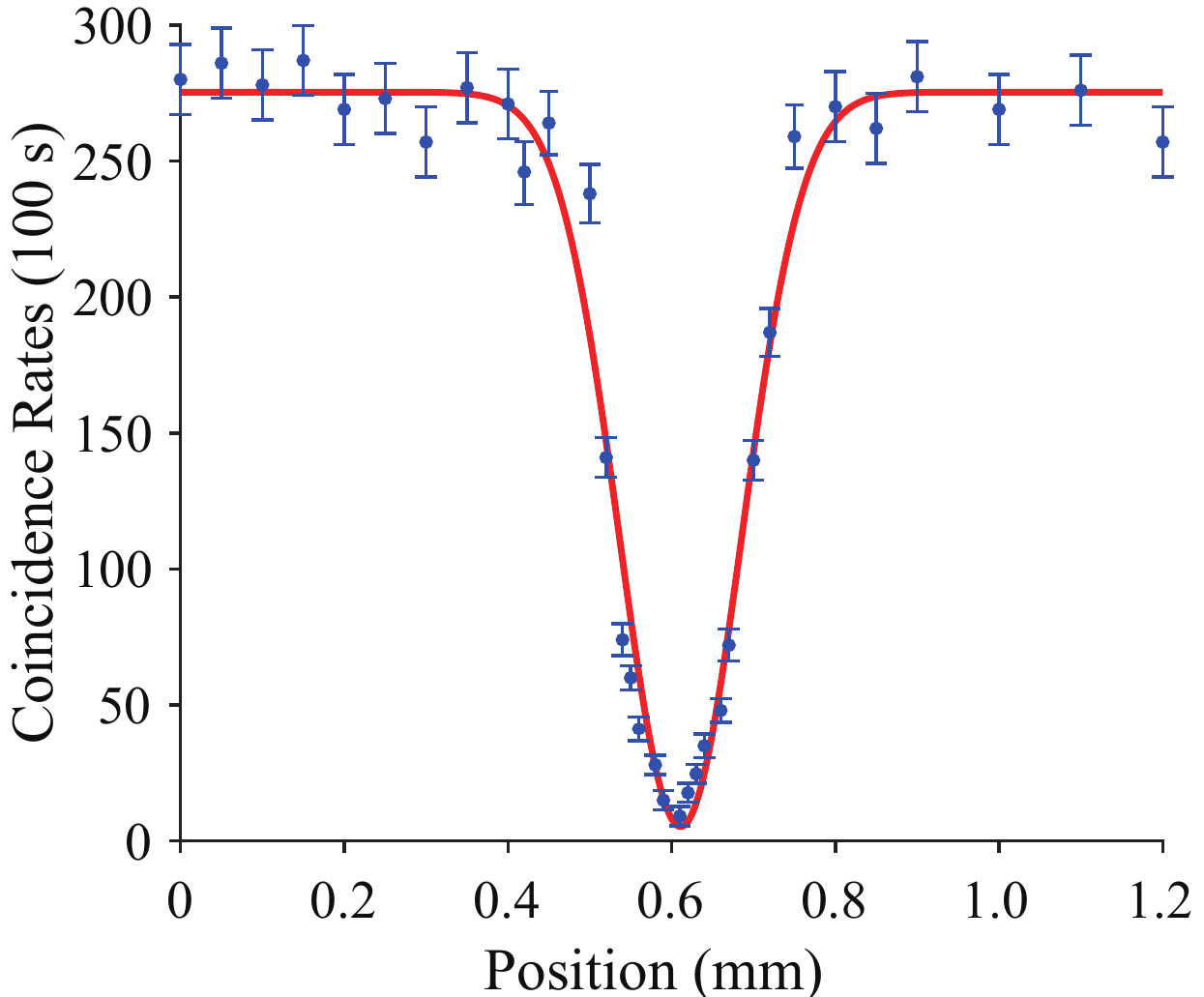}
  \caption{(Color online).(b) The Hong-Ou-Mandel interference of our setup, and its visibility is $0.962\pm0.015$.}\label{fig:hom}
\end{figure}
In our experimental setup, pump pulses pass through two c-BBOs successively. In this way, two photon pairs are prepared in state
\begin{align}
\ket{\Psi_1}&=(\ket{H}_{e_1}\ket{V}_{o_1}-\ket{V}_{e_1}\ket{H}_{o_1})/\sqrt{2}\notag\\
&\otimes(\ket{H}_{e_2}\ket{V}_{o_2}-\ket{V}_{e_2}\ket{H}_{o_2})/\sqrt{2}
\end{align}
with non-vanishing probability, in which subscript 1 and 2 are used to distinguish the photons generated from which c-BBOs. Then by transforming the first photon pair to the Bell state $(\ket{HH}-\ket{VV})/\sqrt{2}$ with a HWP at $45^{\circ}$ for $e_1$ mode photon, we obtained the state
\begin{align}
\ket{\Psi_2}&=(\ket{H}_{e_1}\ket{H}_{o_1}-\ket{V}_{e_1}\ket{V}_{o_1})/\sqrt{2}\notag\\
&\otimes(\ket{H}_{e_2}\ket{V}_{o_2}-\ket{V}_{e_2}\ket{H}_{o_2})/\sqrt{2}.
\end{align}
The vertically polarized photon $\ket{V}_{o2}$ is used as a trigger signal, and we use a HWP at $22.5^{\circ}$ for $e_2$ mode photon.
Afterwards, the two photon in modes $e_1$ and $e_2$ are directly combined on a PBS. The PBS transmits $H$ and reflects $V$, leading to a coincidence registration of a single photon at each output. In this way the two terms $\ket{H}_{e_1}\ket{H}_{o_1}\ket{H}_{e_2}$ and $\ket{V}_{e_1}\ket{V}_{o_1}\ket{V}_{e_2}$ are post-selected, i.e.
\begin{align}
\ket{\Psi_3}=\frac{1}{\sqrt{2}}\left(\ket{H}_{e_1}\ket{H}_{o_1}\ket{H}_{e_2}+\ket{V}_{e_1}\ket{V}_{o_1}\ket{V}_{e_2}\right).
\end{align}
In this step, to achieve great spatial and temporal overlap, a movable prime is used to guarantee the delay between two spatial modes $e_1$ and $e_2$ are equal.
The corresponding Hong-Ou-Mandel interference is illustrated in Fig. \ref{fig:hom}.
We add a QWP at $0^{\circ}$ and a HWP at $22.5^{\circ}$ in each mode $e_1$, $e_2$ and $o_1$ to rotate the state $\ket{\Psi_3}$ to
\begin{align}
\ket{\Psi_4}=\frac{1}{2}{\big(}&\ket{H}_{e_1}\ket{H}_{o_1}\ket{H}_{e_2}-\ket{H}_{e_1}\ket{V}_{o_1}\ket{V}_{e_2}\notag\\
-&\ket{V}_{e_1}\ket{H}_{o_1}\ket{V}_{e_2}-\ket{V}_{e_1}\ket{V}_{o_1}\ket{H}_{e_2}{\big)},
\end{align}
which is the state $\ket{S}$. Additionally, by only changing the QWP's angle from $0^{\circ}$ to  $45^{\circ}$ (red mark), the state $\ket{S'}$ is prepared.

In our experiment, the fidelity $F=(\Tr(\sqrt{\sqrt{\rho_{t}}\rho_{\mathrm{exp}}\sqrt{\rho_{t}}}))^2=0.931\pm0.023$, where $\rho_{\mathrm{exp}}$ and $\rho_{t}=\ket{S}\bra{S}$ are the experimental and theoretical density matrices, respectively. Here, the errors are calculated by assuming Poisson distribution for counting statistics, and re-sampling over recorded data.
In addition, we observed genuine tripartite entanglement of state $\rho_{exp}$ by the tripartite negativity \cite{sabin2008classification},
\begin{align}
N_{123}(\rho)=\left(N_{1,(23)}N_{2,(13)}N_{3,(12)}\right)^{1/3},
\end{align}
where the bipartite negativities are defined as $N_{I,(JK)}=-2\sum_i\sigma_i(\rho^{TI})$, $\sigma_i(\rho^{TI})$ being the negative eigenvalues of $\rho^{TI}$, the partial transpose of $\rho$ with respect to subsystem $I$, $\bra{i_I,j_{JK}}\rho^{TI}\ket{k_I,l_{JK}}=\bra{k_I,l_{JK}}\rho\ket{i_I,j_{JK}}$, with $I=1,2,3$ and $JK=23,13,12$, respectively. 
We calculate the experimental value of the tripartite negativity $N_{123}(\rho_{\mathrm{exp}})$ is $0.974\pm0.013$, which is not equal to the theoretical value $0.9785$, but clearly shows its genuine tripartite entanglement.

\subsection{Experimental results of $\left|S\right>$}
As illustrated in Fig. \ref{fig:exp}, we can use this experiment setup to verify the afore-mention quantum feature for the minimal system.
For the implementation of the three-qubit Hardy-like quantum pigeonhole paradox,
\begin{figure}[t]
\centering
\includegraphics[width=\linewidth]{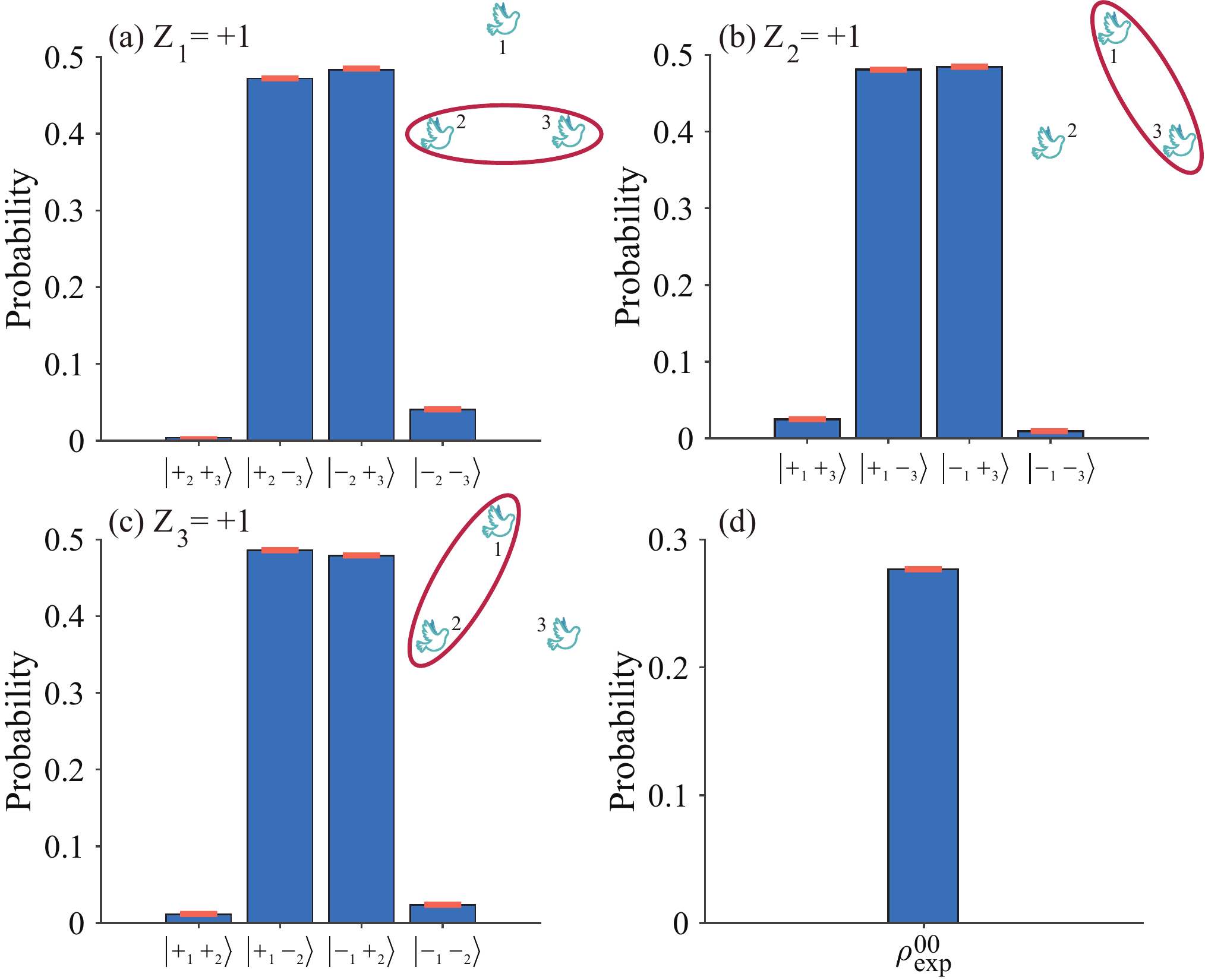}
\caption{(Color online). Data shows the quantum probabilities of measurements for Hardy-like pigeonhole paradox on three-qubit state $\ket{S}$. (a) Experimental results for $X_2X_3$ if $Z_1=+1$; (b) Experimental results for $X_1X_3$ if $Z_2=+1$; (c) Experimental results for $X_1X_2$ if $Z_3=+1$; (d) Experimental results for $Z_1=Z_2=Z_3=+1$. The error bars in both subplots correspond to 1$\sigma$ standard deviation.}
\label{fig:bases}
\end{figure}
we have measured Eqs. (\ref{Hardy-Condi1})-(\ref{Hardy-Condi3}) and (\ref{Hardy-Condi4}). All these experimental results are shown in Fig. \ref{fig:bases}, where
\begin{subequations}
\begin{align}
&\langle X_2X_3\rangle=-0.911\pm0.012,\ \mathrm{if}\ Z_1=+1,\label{exp1}\\
&\langle X_1X_3\rangle=-0.931\pm0.017,\ \mathrm{if}\ Z_2=+1,\label{exp2}\\
&\langle X_1X_2\rangle=-0.930\pm0.020,\ \mathrm{if}\ Z_3=+1,\label{exp3}
\end{align}
\end{subequations}
and the probability of event $Z_1=Z_2=Z_3=+1$ is
\begin{align}\label{exp4}
P(Z_1=Z_2=Z_3=+1)=0.277\pm0.009.
\end{align}
In any local hidden variable theories or noncontextual hidden variable theories, according to the classical pigeonhole principle, events $X_1X_2=X_1X_3=X_2X_3=-1$ cannot jointly hold, which leads to a zero probability of obtaining $Z_1=Z_2=Z_3=+1$. However, our experimental results show a probability of $0.277\pm0.009$ to get $Z_1=Z_2=Z_3=+1$. Although not perfect, it is enough to exhibit the Hardy-type all-versus-nothing phenomenon, and can successfully prove this nonclassical problem.

In fact, the paradox constructed from this state is a representative example corresponding to another kind of PCG (see Fig. \ref{3qubit-PCGs}). Moreover, it is known that there are only two types of Hardy-like quantum pigeonhole paradoxes for three qubit systems. In this sense,  we have completely verified this kind of quantum feature for the minimal systems.

\section{Conclusion and discussion}
In conclusion, with the general projected-coloring graph states, we have studied the general Hardy-like quantum pigeonhole paradoxes, and used the graph projected-coloring problems to portray in which, coexisting post-selections may lead to non-coexisting colorings.
We experimentally verified this feature in the minimal three-qubit systems.
Compared with the original quantum pigeonhole paradox, which needs some sophisticated skills such as weak measurement \cite{chen2019experimental}, the coloring problems (Hardy-like quantum pigeonhole paradoxes) are more feasible in the experimental implementation.

Actually, these quantum paradoxes can be considered as a kind of special many-particle Hardy paradoxes. One can use a special kind of graphs called projected-coloring graphs to design other different paradoxes.
There are more interesting problems which essentially, are a special case of these graph coloring problems, such as quantum magic square problems and quantum map coloring problems in Appendix \ref{app_magic}.
Additionally, the Hardy-like quantum pigeonhole paradox is a powerful tool in studying a kind of GHZ-like nonlocality \cite{Tang2022} unnoticed before. Unlike the common GHZ paradox, the proof for such kind of nonlocality is constructed based on non-perfect correlations rather than constraints induced from stabilizers.

Moreover, since this quantum paradox is essentially a consequence of quantum nonlocality or contextuality, the features of quantum mechanics, not existing in the classical physics, could lead to an operational advantage \cite{matera2016coherent,Hillery2016PRA,Theurer2017super}. Thus one may design some tasks, such as some nonclassical games, multi-party and even high-dimensional untrusted communications, by using this quantum feature as a peculiar quantum resource.

\begin{acknowledgments}
Shihao Ru thanks Zheng-Hao Liu and Yihan Luo for helpful discussions in theory and experiment.
This work at XJTU is supported by the National Nature Science Foundation of China (Grants No. 12074307, No. 11804271, and No. 91736104) and Ministry of Science and Technology of China (Grant No. 2016YFA0301404).
This work at USTC is supported by the National Nature Science Foundation of China (Grants No. 11874345 and No. 11904357), and the Fundamental Research Funds for the Central Universities, USTC Tang Scholarship, Science and Technological Fund of Anhui Province for Outstanding Youth (2008085J02).
\end{acknowledgments}

\appendix

\section{Experimental results of $\left|S^{\prime}\right>$}
\label{app:exp2}
Based on state $\ket{S'}$, we also can construct another Hardy-like quantum pigeonhole paradox. It needs to measure whether the following relations hold
\begin{subequations}
\begin{align}
P(X_2X_3=-1|Z_1=+1)=1,\label{s'1}\\
P(X_1X_3=+1|Z_2=+1)=1,\label{s'2}\\
P(X_1X_2=+1|Z_3=+1)=1,\label{s'3}\\
P(Z_1=Z_2=Z_3=+1)=\frac{1}{4}.\label{s'4}
\end{align}
\end{subequations}
Similarly to $\ket{S}$ in the maintext, we have measured Eqs. (\ref{s'1})-(\ref{s'4}),
\begin{subequations}
\begin{align}
&\left<X_2X_3\right>=-0.921\pm0.011,\ \mathrm{if}\ Z_1=+1,\label{sexp1}\\
&\left<X_1X_3\right>=+0.911\pm0.017,\ \mathrm{if}\ Z_2=+1,\label{sexp2}\\
&\left<X_1X_2\right>=+0.923\pm0.008,\ \mathrm{if}\ Z_3=+1,\label{sexp3}\\
&P(Z_1=Z_2=Z_3=+1)=0.246\pm0.014.\label{sexp4}
\end{align}
\end{subequations}
In any local hidden variable theories or noncontextual hidden variable theories, according to the classical pigeonhole principle, events $X_2X_3=-1$, $X_1X_3=+1$ and $X_1X_2=+1$ cannot jointly hold, which leads to a zero probability of obtaining $Z_1=Z_2=Z_3=+1$. But our experimental results show a probability of $0.246\pm0.014$ to get $Z_1=Z_2=Z_3=+1$. All these experimental results, including $\ket{S}$ and $\ket{S'}$, can discard the Hardy-like all-versus-nothing phenomenon, and successfully verify the Hardy-like quantum pigeonhole paradoxes.

\section{Three kinds of PCG states} \label{app_states}
We here enumerate three kinds of general quantum states that can be used for Hardy-like quantum pigeonhole paradox once again. The simplest quantum state $\ket{S}$ in each category is consistent.
The first kind of quantum states for Hardy-like pigeonhole paradox are
\begin{align}
\ket{S_1(n)}&=\frac{1}{\sqrt{n+1}}\left[\ket{\underbrace{0\cdots00}_n}-\ket{W;n,1}\right],\label{S1-}\\
\ket{W;n,k}&=\ket{\underbrace{k\cdots kk}_{n-1}0}+\ket{\underbrace{k\cdots k}_{n-2}0k}+\cdots\notag\\
&+\ket{k0\underbrace{k\cdots k}_{n-2}}+\ket{0\underbrace{kk\cdots k}_{n-1}},
\end{align}
where $n\ge 3$ and $n$ is odd. Apart from quantum states $\ket{S_1}$, another two kinds of quantum states are $n$-qubits states and $(d+1)$-qudit states, that can be written as
\begin{align}
\ket{S_2}=\frac{1}{\sqrt{1+C_n^2}}\left[\ket{\underbrace{0\cdots00}_n}-\ket{\Phi}\right],\label{S2}\\
\ket{S_3}=\frac{1}{d}\left[\ket{\underbrace{0\cdots00}_{d+1}}-\sum_{l=1}^{d-1}e^{i l\theta}\ket{W;d+1,l}\right],\label{S3}
\end{align}
respectively.
Here $\ket{\Phi}=\sum_{perm}\ket{00\underbrace{11\cdots11}_{n-2}}$, where the summation is over all permutations of $\ket{00\underbrace{11\cdots11}_{n-2}}$, $n\ge 3$ and $n$ is odd.

\label{app:multiple_qubits_}
\subsection{Multi-qubit case 1}
For quantum states $\ket{S_1(n)}$, we can have
\begin{align}\label{eq2}
X_2X_3\cdots X_n&=-1,\ \mathrm{if}\ Z_1=1;\notag\\
X_1X_3\cdots X_n&=-1,\ \mathrm{if}\ Z_2=1;\notag\\
&\cdots\notag\\
X_1X_2\cdots X_{n-1}&=-1,\ \mathrm{if}\ Z_n=1.
\end{align}
According to its classical assignment assumption in HVTs, $X_l$ and $Z_l$ can be assigned predefined values $x_l$ and $z_l$, respectively, where $v=\pm1,v\in\{z_l,x_l\}$.
It follows that if the observables from one context (mutually commuting) satisfy a certain algebraic relationship, then the assigned predefined values obey the same algebraic constraint, i.e.,
\begin{align}\label{eqC}
x_2x_3\cdots x_n&=-1,\ \mathrm{if}\ z_1=1;\notag\\
x_1x_3\cdots x_n&=-1,\ \mathrm{if}\ z_2=1;\notag\\
&\cdots\notag\\
x_1x_2\cdots x_{n-1}&=-1,\ \mathrm{if}\ z_n=1.
\end{align}
Using the form of conditional event, these $n$ events can be expressed as $\{x_2x_3\cdots x_n=-1|z_1=+1\}$, $\{x_1x_3\cdots x_n=-1|z_2=+1\}$, $\cdots$, and $\{x_1x_2\cdots x_{n-1}=-1|z_n=+1\}$.
The probabilities of those $n$ events are
\begin{align}\label{conditional_form}
&P_C(x_2x_3\cdots x_n=-1|z_1=1)\notag\\
&=P_C(x_1x_3\cdots x_n=-1|z_2=1)=\cdots\notag\\
&=P_C(x_1x_2\cdots x_{n-1}=-1|z_n=1)=1,
\end{align}
in which we utilize subscript $C(Q)$ to denote classical (quantum) probability.

According the probabilities being $1$ of those $n$ events, if event $z_1=z_2=\cdots=z_n=1$ occurs, event $x_2x_3\cdots x_n=x_1x_3\cdots x_n=x_1x_2\cdots x_{n-1}=-1$ will also occur at the same time and be with the same probability, hence we can get
\begin{align}
(x_1x_2\cdots x_{n-1}x_n)^{n}=-1.
\end{align}
This method, multiplying both sides of equations, usually be used in the all-versus-nothing proof.
But obviously, $(x_1x_2\cdots x_{n-1}x_n)^{n-1}$ must be nonnegative since all $x_l$ are real numbers and $n-1$ is even. That is the event $x_2x_3\cdots x_n=x_1x_3\cdots x_n=x_1x_2\cdots x_{n-1}=-1$ cannot occur via a classical assignment method.
Therefore, it is logically appropriate that events $z_1=z_2=\cdots=z_n=1$ and $x_2x_3\cdots x_n=x_1x_3\cdots x_n=x_1x_2\cdots x_{n-1}=-1$ occur simultaneously only when
\begin{align}\label{eq5}
P_C(z_1=z_2=\cdots=z_n=1)=0.
\end{align}
On the other hand, the probability of classical event $z_1=z_2=\cdots=z_n=1$ is equal to its quantum scenery
\begin{align}
\ P_Q(Z_1=Z_2=\cdots=Z_n=1)=\frac{1}{n+1}.\label{eq4}
\end{align}
Eqs. (\ref{eq5}) and (\ref{eq4}) are contradictory. It is indicated that we cannot make classical assignments to observable values, and the hypothesis of HVTs is invalid.
Therefore, for testing the paradox based on the quantum state $\ket{S_1}$, we only need to measure Eqs. (\ref{eq2}) and (\ref{eq4}).

If $n=3$, the state is
\begin{align}
  \ket{S_1(n=3)}=(\ket{000}-\ket{011}-\ket{101}-\ket{110})/2,
\end{align}
which is the same as $\ket{S}$.

\subsection{Multi-qubit Case 2} 
For quantum states $\ket{S_2(n)}$, we can have
\begin{align}
X_mX_n=-1,\ \mathrm{if\ all}\ Z_{j\neq m,n}=1\ \mathrm{and}\ m\neq n.
\end{align}
Similar to above, according to its classical assignment assumption in HVTs, its assigned predefined values obey the same algebraic constraint, i.e.
\begin{align}
x_mx_n=-1,\ \mathrm{if\ all}\ z_{j\neq m,n}=1\ \mathrm{and}\ m\neq n.
\end{align}
Using the form of conditional event, the probabilities of these $C_n^2$ events are
\begin{align}
&P_C(x_1x_2=-1|z_3=\cdots=z_n=1)\notag\\
&=P_C(x_1x_3=-1|z_2=z_4\cdots=z_n=1)=\cdots\notag\\
&=P_C(x_{n-1}x_n|z_1=z_2\cdots=z_n=1)=1.
\end{align}
Hence we can also get
\begin{align}
(x_1x_2\cdots x_{n-1}x_n)^{n-1}=-1.
\end{align}
But obviously, $(x_1x_2\cdots x_{n-1}x_n)^{n-1}$ must be nonnegative since all $x_l$ are $+1$ or $-1$ and $n-1$ is even.
Therefore, it is logically appropriate that events $z_1=z_2=\cdots=z_n=1$ and $x_1x_2=\cdots=x_1x_n=\cdots=x_{n-1}x_n=-1$ occur simultaneously only when
\begin{align}\label{eq18}
P_C(z_1=z_2=\cdots=z_n=1)=0.
\end{align}
On the other hand, the probability of classical event $z_1=z_2=\cdots=z_n=1$ is equal to its quantum scenery
\begin{align}
\ P_Q(Z_1=Z_2=\cdots=Z_n=1)=\frac{2}{n(n-1)+2}.\label{eq19}
\end{align}
Eqs. (\ref{eq18}) and (\ref{eq19}) are contradictory. It is indicated that we cannot make classical assignments to observable values, and the hypothesis of HVTs is invalid.
And if $n=3$, the state is
\begin{align}
  \ket{S_2(n=3)}=(\ket{000}-\ket{011}-\ket{101}-\ket{110})/2,
\end{align}
which is the same as $\ket{S}$.

\subsection{Multi-qudit Case} 
\label{sub:multiple_qudits}
The Pauli matrix group has applications in quantum computation, quantum teleportation, and other quantum protocols. This group is defined for a single qudit in the following manner,
\begin{align}
Z_d=&\sum_{n=0}^{d-1}e^{\mi n\theta}\ket{n}\bra{n},\\
X_d=&\sum_{n=0}^{d-1}\ket{n\oplus1}\bra{n},
\end{align}
where $\theta=2\pi/d$, $a\oplus b =(a+b)\ \mathrm{mod}\ d$, and $XZ=ZXe^{i\theta}$. Furthermore, the $Y$ gate can be written $Y=XZ$.

For quantum states $\ket{S_3(n)}$, we can have
\begin{align}
\prod_{j=1,j\neq k}^{d+1}X_j=e^{-i\theta},\ \mathrm{if}\ Z_k=1,
\end{align}
where $k=1,2,\cdots,d+1$. According to its classical assignment assumption in HVTs, its assigned predefined values obey the same algebraic constraint, i.e.
\begin{align}
\prod_{j=1,j\neq k}^{d+1}x_j=e^{-i\theta},\ \mathrm{if}\ z_k=1.
\end{align}
Using the form of conditional event, the probabilities of these $d$ events are
\begin{align}\label{conditional_form3}
&P_C(x_2x_3\cdots x_{d+1}=e^{-i\theta}|z_1=1)\notag\\
&=P_C(x_1x_3\cdots x_{d+1}=e^{-i\theta}|z_2=1)=\cdots\notag\\
&=P_C(x_1x_2\cdots x_{d}=e^{-i\theta}|z_{d+1}=1)=1,
\end{align}
According the probabilities being $1$ of those $n$ events, if event $z_1=z_2=\cdots=z_d=z_{d+1}=1$ occurs, event $x_2x_3\cdots x_{d+1}=x_1x_3\cdots x_{d+1}=\cdots=x_1x_2\cdots x_{d}=e^{-i\theta}$ will also occur at the same time and be with the same probability, hence we can get
\begin{align}
(x_1x_2\cdots x_dx_{d+1})^{d}=e^{-id\theta}=-1.
\end{align}
But obviously, $(x_1x_2\cdots x_{n-1}x_n)^{n-1}$ must be nonnegative.
Therefore, it is logically appropriate that events $z_1=z_2=\cdots=z_d=z_{d+1}=1$ and $x_2x_3\cdots x_{d+1}=x_1x_3\cdots x_{d+1}=\cdots=x_1x_2\cdots x_{d}=e^{-i\theta}$ occur simultaneously only when
\begin{align}\label{eq28}
P_C(z_1=z_2=\cdots=z_d=z_{d+1}=1)=0.
\end{align}
On the other hand, the probability of classical event $z_1=z_2=\cdots=z_d=z_{d+1}=1$ is equal to its quantum scenery
\begin{align}\label{eq29}
P_Q(z_1=z_2=\cdots=z_d=z_{d+1}=1)=\frac{1}{d^2}.
\end{align}
Eqs. (\ref{eq28}) and (\ref{eq29}) are contradictory. It is indicated that we cannot make classical assignments to observable values, and the hypothesis of HVTs is invalid.
And if $d=2$, the state is
\begin{align}
  \ket{S_3(d=2)}=(\ket{000}-\ket{011}-\ket{101}-\ket{110})/2,
\end{align}
which is also the same as $\ket{S}$.

\section{Quantum magic square paradox}\label{app_magic}
Putting $1$ to $n^2$ ($n\geq3$) in an $n\times n$ table such that the sum of the numbers in each row, column, and both diagonals are the same, one can get an $n$-th order magic square.
A simplified version is the binary case in which only 0 and 1 are allowed to be filled, and such magic squares are called binary magic squares.
Similar to the construction of PCGs,  if mapping 0 and 1 to $X=1$ and $X=-1$ respectively, one can build a kind of quantum magic squares and the associated Hardy-like quantum magic square paradox.

We present a Hardy-like quantum magic square paradox based on a second order quantum magic square as an example, seen in Supplementary Fig. \ref{magicsquare and four colors}(a).
Four qubits are put in the $2\times 2$ magic square and labeled by 1, 2, 3 and 4 from left to right and from top to bottom.
As referred above, the predefined value of $X_i$ on the $i$-th qubit mapped to $0$ or $1$ can be considered as the associated binary number filled in the magic square.

Consider the state $|M\rangle=(|0000\rangle-|1100\rangle-|1010\rangle-|1001\rangle-|0110\rangle-|0101\rangle-|0011\rangle)/\sqrt{7}$, one can check that
\begin{subequations}\label{quantum-magic-square}
\begin{align}
  P(X_3X_4=-1|Z_1=Z_2=1) &= 1,\label{quantum-magic-square1} \\
  P(X_2X_4=-1|Z_1=Z_3=1) &= 1,\label{quantum-magic-square2}\\
  P(X_2X_3=-1|Z_1=Z_4=1) &= 1,\label{quantum-magic-square3}\\
  P(X_1X_4=-1|Z_2=Z_3=1) &= 1, \label{quantum-magic-square4}\\
  P(X_1X_3=-1|Z_2=Z_4=1) &= 1, \label{quantum-magic-square5}\\
  P(X_1X_2=-1|Z_3=Z_4=1) &= 1, \label{quantum-magic-square6}\\
  P(Z_1=Z_2=Z_3=Z_4=1) &= \frac{1}{7}.\label{quantum-magic-square7}
\end{align}
\end{subequations}
Therefore, if one measure $Z_1$ and $Z_2$ on qubits 1 and 2 and get the outcomes of $Z_1=1$ and $Z_2=1$, one can infer that $X_3X_4=-1$ and likewise the others. If  $Z_1=Z_2=Z_3=Z_4=1$ can jointly hold with a nonzero probability, then in the classical framework, $X_1X_2=X_1X_3=X_1X_4=X_2X_3=X_2X_4=X_3X_4=-1$ is jointly hold with a nonzero probability. Using a similar argument of the case of three qubit Hardy-like quantum pigeonhole paradox, one can see that is impossible. Then we have constructed a Hardy-like quantum magic square paradox.

\begin{figure}[h]
\includegraphics[width=0.25\linewidth]{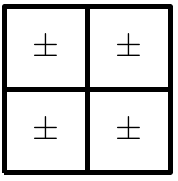}
\caption{A $2\times2$ quantum magic square occupied by a four-qubit quantum state.}\label{magicsquare and four colors}
\end{figure}

The following examples are more general Hardy-like quantum magic square paradoxes.

{\it Example 1}.--- Consider a $(4\times 4)$-qubit PCG state $|M(16)\rangle=\frac{1}{\sqrt{11}}(|\vec{0}\rangle_{\mS}-\sum_{i=1}^{10}|\vec{1}\rangle_{\mS_{i}}|\vec{0}\rangle_{\bar{\mS}_i})$, where $\mS=\{1,2,\cdots,16\}$ and $\{\mS_i|i=1,2,\cdots,10\}=\{\{4k+1,4k+2,4k+3,4k+4\}|k=0,1,2,3\}\cup\{\{l,4+l,8+l,12+l\}|l=1,2,3,4\}
\cup\{1,6,11,16\}\cup\{4,7,10,13\}$.
Assume that $|M(16)\rangle$ can be modeled by LHV. Consider a run of the experiment for which $Z_1,Z_2,\cdots,Z_{16}$  are measured and the results $Z_1=Z_2=\cdots=Z_{16}=1$ are obtained. Similar to the argument of the  HLQP paradox in the main text, one can finally conclude that $\prod_{j\in\mS_1}X_j=\prod_{j\in\mS_2}X_j=\cdots=\prod_{j\in\mS_{10}}X_j=-1$. Based on that,  one can find some solutions for $X_1,X_2,\cdots,X_{16}$. There is no contradiction.

Notice that $\prod_{i=1}^{10}(\prod_{j\in\mS_i}X_j)=X_1X_4X_6X_7X_{10}X_{11}X_{13}X_{16}=1$. One can consider another PCG state $|\tilde{M}(16)\rangle=\frac{1}{\sqrt{12}}(\sqrt{11}|M(16)\rangle-|1001011001101001\rangle)$. Namely, we impose a new conditional constraint:
If $Z_2=Z_3=Z_5=Z_8=Z_9=Z_{12}=Z_{14}=Z_{15}=1$ are obtained, then necessarily $X_1X_4X_6X_7X_{10}X_{11}X_{13}X_{16}=-1$. Next, we also consider a run of the experiment for which $Z_1,Z_2,\cdots,Z_{16}$  are measured and the results $Z_1=Z_2=\cdots=Z_{16}=1$ are obtained. Then this extra constraint ensures that there is no consistent solution for $X_1,X_2,\cdots,X_{16}$ in classical world according to pigeonhole principle.

Let $m_r=(X_r+1)/2$ be the number arranged in the $r$-th grid of the binary magic square.
Notice that $\prod_{j\in\mS_i}X_{j}=(-1)^{\oplus_{j\in\mS_i}m_j}=-1$ and
$X_1X_4X_6X_7X_{10}X_{11}X_{13}X_{16}=(-1)^{m_1\oplus m_4\oplus m_6\oplus m_7\oplus m_{10}\oplus m_{11}\oplus m_{13}\oplus m_{16}}=-1$. Note that here $\oplus$  stands for addition modulo 2 rather than modulo $n$.
It follows that $\oplus_{j\in\mS_1}m_j=\oplus_{j\in\mS_2}m_j=\cdots=\oplus_{j\in\mS_{10}}m_j=m_1\oplus m_4\oplus m_6\oplus m_7\oplus m_{10}\oplus m_{11}\oplus m_{13}\oplus m_{16}=1$, a contradiction(the assumption of local realism can ``induce" a binary magic square which is forbidden in classical world). Then we can get a $4$-order conditional(with an extra constraint) Hardy-like quantum magic square paradox.

{\it Remarks.}--- Commonly, there are some prescribed constraints for a classical magic square(e.g. the $3\times3$ conventional magic square is arranged with numbers $1,2,\cdots,9$). Even for a binary magic square, usually the number of zeros(or ones) to be arranged should be prescribed(e.g. $4$ zeroes and $5$ ones for a $3\times3$ binary magic square). However, for a quantum binary magic square, we would like to choose some other constraints, such as the extra constraint imposed in the above example. After all, our goal is just to show that a classically impossible magic square might be probabilistically produced if the associated quantum state admits a LHV model.

{\it Example 2}.--- Consider a nine-qubit PCG state $|M(9)\rangle=\frac{1}{3}(|000000000\rangle-|111000000\rangle-|0001110000\rangle-|000000111\rangle
-|100100100\rangle-|010010010\rangle-|001001001\rangle-|100010001\rangle-|001010100\rangle)$.
Assume that $|M(9)\rangle$ can be modeled by LHV. Consider a run of the experiment for which $Z_1,Z_2,Z_3,Z_4,Z_5,Z_6,Z_7,Z_8,Z_9$  are measured and the results $Z_1=Z_2=Z_3=Z_4=Z_5=Z_6=Z_7=Z_8=Z_9=1$ are obtained. Likewise, one can conclude that the relations $X_1X_2X_3=X_4X_5X_6=X_7X_8X_9=X_1X_4X_7=X_2X_5X_8=X_3X_6X_9=X_1X_5X_9=X_3X_5X_7=-1$ should be satisfied. There is also no contradiction.

Notice that the  product of these above eight relations gives rise to $X_1X_3X_7X_9=1$. One can use another PCG state $|\tilde{M}(9)\rangle=\frac{1}{\sqrt{10}}(3|M(9)\rangle-|101000101\rangle)$ to construct a HLQP paradox. Likewise,  consider a run of the experiment for which $Z_1,Z_2,Z_3,Z_4,Z_5,Z_6,Z_7,Z_8,Z_9$  are measured and the results $Z_1=Z_2=Z_3=Z_4=Z_5=Z_6=Z_7=Z_8=Z_9=1$ are obtained. Besides $X_1X_2X_3=X_4X_5X_6=X_7X_8X_9=X_1X_4X_7=X_2X_5X_8=X_3X_6X_9=X_1X_5X_9=X_3X_5X_7=-1$, one can get an extra relation $X_1X_3X_7X_9=-1$.
All such relations contradict with pigeonhole principle.

Let $m_k=(X_k+1)/2$ be the number arranged in the $k$-th grid of the binary magic square. One can get
$m_1\oplus m_2\oplus m_3=m_4\oplus m_5\oplus m_6=m_7\oplus m_8\oplus m_9=m_1\oplus m_4\oplus m_7=m_2\oplus m_5\oplus m_8=m_3\oplus m_6\oplus m_9=m_1\oplus m_5\oplus m_9=m_3\oplus m_5\oplus m_7=1$ and $m_1\oplus m_3\oplus m_7\oplus m_9=1$, which cannot hold simultaneously in classical world.  This contradiction can induce another conditional Hardy-like quantum magic square paradox.

Note: One can also consider the case that a magic square stays on a torus(e.g. Fig.2-(e) in the main text), and one may construct a similar Hardy-like quantum magic square paradox(under some extra constraints).

{\it Example 3}.--- We generalize the notion of binary magic square to the $n$-dimensional case. For example, a $3$-dimensional binary magic square of order 2 is  an arrangement of $k$ ones and $2^3-k$ zeros in a $2\times2\times2$-cube, such that the XOR sum of the numbers  in each edge, four main diagonals, and twelve other diagonals is the same.

Consider an eight-qubit PCG state $|M(8)\rangle=\frac{1}{\sqrt{C_8^2+1}}(|00000000\rangle-\sum_{i=1}^{C_8^2}|11\rangle_{\mS_i}|000000\rangle_{\bar{\mS}_i})$, where $\mS_i=\{a_i,b_i\}$ and $a_i\neq b_i\in\{1,2,3,\cdots,8\}$. Also assume that $|M(8)\rangle$ can be modeled by LHV. Consider a run of the experiment for which $Z_1,Z_2,\cdots,Z_{8}$  are measured and the results $Z_1=Z_2=\cdots=Z_{8}=1$ are obtained. Likewise, one can finally conclude that $\prod_{j\in\mS_1}X_j=\prod_{j\in\mS_2}X_j=\cdots=\prod_{j\in\mS_{28}}X_j=-1$, which contradict with pigeonhole principle.

Let $m_r=(X_r+1)/2$.  Notice that $\prod_{j\in\mS_i}X_{j}=(-1)^{\oplus_{j\in\mS_i}m_j}=-1$. Then one can get
$\oplus_{j\in\mS_1}m_j=\oplus_{j\in\mS_2}m_j=\cdots=\oplus_{j\in\mS_{28}}m_j=1$, a contradiction. Namely, we get a generalized Hardy-like quantum magic square paradox.

\nocite{*}


%

\end{document}